\documentclass{article}
\usepackage{emulateapj}

\usepackage{psfig}

\def\ledd{L_{\rm Edd}}

\def\msun{{\,M_\odot}}

\newcommand\sch{Schwarzschild~}
\newcommand\fe{Fe K$\alpha$\ }
\newcommand\rozanska{R\'o$\dot{\rm z}$a\'nska }
\newcommand\zycki{$\dot{\rm Z}$ycki }

\newcommand\fx{F_{\rm x}}
\newcommand\lx{L_{\rm x}}
\newcommand\trec{t_{\rm rec}}
\newcommand\thydro{t_{\rm h}}
\newcommand\ttherm{t_{\rm th}}
\newcommand\dm{\dot{m}}
\newcommand\fdisk{F_{\rm disk}}
\newcommand\ldisk{L_{\rm disk}}

\newcommand\tth{t_{\rm th}}
\newcommand\fline{F_{\rm line}}
\newcommand\pgas{P_{\rm gas}}
\newcommand\pcrit{P_{\rm crit}}

\def\>{$>$}
\def\<{$<$}

\def\simlt{\lower.5ex\hbox{$\; \buildrel < \over \sim \;$}}
\def\simgt{\lower.5ex\hbox{$\; \buildrel > \over \sim \;$}}
\def\sqr#1#2{{\vcenter{\hrule height.#2pt
      \hbox{\vrule width.#2pt height#1pt \kern#1pt
         \vrule width.#2pt}
      \hrule height.#2pt}}}

\begin{document}

\title{On time-dependent X-ray reflection by photoionized accretion
disks: implication for Fe K$\alpha$ line reverberation studies of AGN}

\author{Sergei Nayakshin\altaffilmark{1} and Demosthenes Kazanas}
\affil{NASA/GSFC, LHEA, Code 661, Greenbelt, MD, 20771}
\altaffiltext{1}{Also Universities Space Research Association}

\begin{abstract}
We perform a first study of time-dependent X-ray reflection in
photo-ionized accretion disks. We assume a step-functional change in
the X-ray flux and use a simplified prescription to describe the time
evolution of the illuminated gas density profile in response to
changes in the flux. We find that the dynamical time for re-adjustment
of the hydrostatic balance is an important relaxation time scale of
the problem since it affects evolution of the ionization state of the
reflector.  Because of this the \fe line emissivity depends on the
shape and intensity of the illuminating flux in {\em prior} times, and
hence it is {\em not} a function of the instantaneous illuminating
spectrum. Moreover, during the transition, a prominent Helium-like
component of the \fe line may appear. As a result, the \fe line flux
may appear to be completely uncorrelated with X-ray continuum flux on
time scales shorter than the dynamical time.  In addition, the
time-dependence of the illuminating flux may leave imprints even on
the time-averaged \fe line spectra, which may be used as an additional
test of accretion disk geometry.  Our findings appear to be important
for the proposed \fe line reverberation studies {\em in lamppost-like
geometries} for accretion rates exceeding about $\sim 1\%$ of the
Eddington value.  However, most AGN do not show Helium-like lines that
are prominent in such models, probably indicating that these models
are not applicable to real sources.
\end{abstract}

\keywords{accretion, accretion disks ---radiative transfer --- line:
formation --- X-rays: general}

\section{Introduction}\label{sect:intro}


Broad Fe K$\alpha$ line emission and the so-called reflection ``hump"
centered around $\sim 30$ keV are thought to be significant
observational signatures of the presence of cold matter in the
vicinity of the event horizons of accreting black holes in Active
Galactic Nuclei (AGN) and Galactic Black Hole Candidates (GBHC). These
spectral features result from the X-ray illumination of the surface of
optically thick and relatively cold matter (e.g., Basko, Sunyaev \&
Titarchuk 1974; Lightman \& White 1988; White, Lightman \& Zdziarski
1988; George \& Fabian 1991; Magdziarz \& Zdziarski 1995; Poutanen,
Nagendra \& Svensson 1996). Because the accretion disks which power
this class of sources are optically thick and relatively cool they
were considered to be the natural sites at which these spectral
features are produced; futhermore, it has been suggested that accurate
measurements of the properties of these features can be used to
determine the properties of the associated accretion disks. For
example, the normalization of the reflection component can provide an
estimate of the solid angle covered by the accretion disk as seen from
the X-ray source, while the precise \fe line profile constrains the
radial structure of such a disk and the underlying space-time geometry
(e.g., Fabian et al. 1989).

Unfortunately, the {\em steady-state} line profile depends on many
parameters -- e.g., the viewing angle, the inner and outer disk radii,
and the \fe emissivity as a function of radius. It is impossible to
constrain the actual values of all of these parameters simultaneously
(see, e.g., Nandra et al. 1999 for the example of NCG~3516). Moreover,
the line profile itself cannot constrain the mass of the black
hole. For this reason, in the absence of telescopes of sufficiently
high angular resolution, it is hoped that the degeneracy in the
steady-state models can be removed through observations of time
variability of the spectral features (Stella 1990, Matt \& Perola
1992; Campana \& Stella 1993, 1995). Complimentary, this would also
set limits on the physical size of the emitting region and hence
determine the black hole mass. Reynolds et al. (1999) and Young \&
Reynolds (2000) suggested that a red-ward moving ``bump" in the \fe
line profile after an X-ray flare would be a robust signature of a
maximally rotating Kerr black hole, thus providing information about
the black hole properties inaccessible by other means; it is also
hoped that similar observations may even be used to test General
Relativity in the strong limit (see also Ruszkowski 2000).

Observational facts gathered to date signal that correlated spectral
and timing analysis of data should indeed be very fruitful in testing
models: almost every time that theories were subjected to
observational scrutiny, AGN produced surprizes that required
modifications of these theories. For example, in the completely
analogous situation of optical and UV AGN line emission, reverbaration
mapping campaigns indicated that the size of the Broad Line Region
(BLR) is smaller by roughly a factor of ten than previously thought
(see Netzer \& Peterson 1997 for a review). Correlated Optical -- UV
(OUV) continuum variability observations have shown the lags between
these two bands to be much shorter than those expected on the basis of
the prevailing accretion disk models. This led to the (reasonable)
suggestion that the correlated O-UV variability ought to result from
the reprocessing of X-rays onto the O-UV emitting disks (Krolik et
al. 1991), only to be challenged later by correlated OUV-X-ray
monitoring campaigns (Nandra et al. 1998; Edelson et al. 2000) that
did not show the expected correlations.

A few existing attempts to observe \fe line reverberation also
provided frustrating results. A search for the \fe line response to
continuum variations was performed for MCG--6--30--15 (Lee et
al. 2000; Reynolds 2000) and NGC 5548 (Chiang et al. 2000). Whereas
the continuum X-ray flux was strongly variable on short time scales,
the \fe line flux did not correlate with it. Vaughan \& Edelson (2001)
have split the light curve of MCG-6-30-15 into much shorter time
intervals than those done by the previous authors -- in bins of RXTE
orbital sampling period. While \fe line flux is found to vary on very
short time scales, it is not correlated with the continuum X-ray flux.
This behavior is not predicted by any of the aforementioned
theoretical papers, and it clearly deserves an explanation.

The purpose of this paper is to study the microphysics of the problem
-- i.e., the time-dependent photo-ionized X-ray reflection -- in
greater detail than it has been done so far. The existing theoretical
literature on \fe line reverberation (Reynolds et al. 1999; Young \&
Reynolds 2000; Ruszkowski 2000) are based on the assumption that the
illuminated gas maintains a constant density, independent of the value
of the X-ray flux, a common assumption of pre-year-2000 literature
(e.g., Ross \& Fabian 1993; Matt, Fabian \& Ross 1993, 1996; \zycki et
al. 1994; Ross, Fabian \& Brandt 1996). In this case one is fortunate
in that it is possible to formulate a set of rules that determine the
emissivity of the line as a function of the {\em instantaneous} value
of the ionizing flux, a fact which greatly simplifies this extremely
complex problem.

However, several other treatments of the static X-ray reflection
problem did not use the constant density assumption considering it may
be too restrictive or unrealistic (e.g., Basko, Sunyaev \& Titarchuk
1974; Raymond 1993; Ko \& Kallman 1994; \rozanska \& Czerny 1996;
Ross, Fabian \& Young 1999; Nayakshin, Kazanas \& Kallman 2000, NKK
hereafter). These authors employed the condition of hydrostatic
balance in order to compute (rather than assume) the gas density. The
code of NKK was designed for conditions typical of the inner regions
of AGN accretion disks and is especially well placed to compare the
predictions of these models with those employing constant density.
Substantial differences in the reflected spectra were found (see also
Ballantyne, Ross \& Fabian 2001). The root of these differences can be
traced to the presence of a thermal ionization instability (TII) under
conditions of constant pressure, previously known from studies of AGN
line emitting clouds (e.g., Krolik, McKee \& Tarter 1981; see also
Field 1965). Because this instability is present only under constant
pressure conditions, it is absent when the constant density assumption
is imposed. NKK, Done \& Nayakshin (2001) and Ballantyne et al. (2001)
conclude that the constant density models are not to be trusted.
Therefore, the {\em quantitative} \fe line reverberation pattern may
in fact be different from that found by Reynolds et al. (1999), Young
\& Reynolds (2000) and Ruszkowski (2000), and it is the purpose of
this paper to attempt to clarify in which way. We must mention from
the outset, however, that because of the extreme computational
overhead, we only study local reflection spectra and do not integrate
over the entyre disk surface to produce full time-dependent responce
of \fe line. Our study is thus only an initial step in the right
direction and an additional work will be required to calculate the
final \fe line transfer function.

The main new effect that we find below can be summarized as follows.
Within the constant density approach, the ionization state of the
illuminated gas is determined mainly by the instantaneous value of the
ionization parameter $\xi= 4\pi \fx/n$, where $\fx$ is the the X-ray
flux and $n$ is the local gas density. Given that $n$ is assumed to be
unchanged, the resulting \fe line flux is a function of $\fx$
only. All the line reverberation studies to date have made use of this
fact (e.g., see \S IIa in Blandford \& McKee 1982 and \S 2 in Reynolds
et al. 1999).

In the context of hydrostatic balance models, the gas density at each
point is calculated from the hydrostatic equilibrium condition
(together with ionization and energy balances, and radiation
transfer). A change in the X-ray flux will lead to changes in the gas
temperature profile, disturbing hydrostatic balance. The latter can be
re-established after time $t$ of order of the local dynamical
time. During this time the ionized disk structure is out of
equilibrium with the incident X-ray flux, and hence the \fe line
emissivity is not a function of $\fx$. In other words, the hydrostatic
balance models have a relaxation time scale that is not present in the
constant density models, and it is important to understand possible
implications of this for the general X-ray reflection problem.

The structure of the paper is as follows. In \S 2 we establish the
important physical quantities and observables that determine the {\em
static} reflected spectrum in the general case of a photo-ionized
reflector, and then discuss which of these can realistically be
measured in future \fe line reverberation studies. In \S 3 we estimate
time scales of the problem, while in \S 4 numerical approach and tests
are presented. Discussion of the results and their implications for
current observations are given in \S 5, and \S 6 lists our
conclusions.

\section{On static X-ray reflection}\label{sect:static}

X-ray emission is a ubiquitous feature of the AGN spectra.
It is believed that the X-rays are emitted in a hot corona
which is located in the vicinity of the compact object, 
just as the cooler, optically thin, geometrically thick 
accretion disks. The proximity of these two distinct 
components of AGN emission implies that accretion disks are 
expected to be strongly photo-ionized by the X-rays, the more so
the larger the accretion rate is (e.g., Matt et al. 1993, 1996,
Nayakshin 2000b). Based on the extensive literature and
our own calculations, we find that the following quantities determine
the spectrum of the reprocessed X-rays:

(1) The absolute value of the incident X-ray flux around the iron
recombination band, $\fx$; this is the most important parameter 
associated with the \fe line emission.  In the
case of neutral reflection, the line emissivity, $\fline \propto \fx$,
and therefore the line EW is constant (unless the spectrum of 
the reprocessed X-rays is variable).

(2) The value of the locally produced thermal disk emission,
$\fdisk$. The absolute value of ratio $\fx/\fdisk$ affects the
temperature of the ionized skin (see Nayakshin \& Kallman 2001).

(3) The photon spectral index of the incident radiation,
$\Gamma$. George \& Fabian (1991) have shown that the EW of the line
increases by about a factor of 1.6 as $\Gamma$ decreases from 2.3 to
1.3. For neutral reflection, the physics of this dependence is rather
simple -- as $\Gamma$ changes, the amount of X-ray photons capable of
photo-ionizing the Fe K-shell changes, and so does the EW of the line. 

In the case of photo-ionized reflection, the spectral index  plays a 
more prominent role as it also determines the temperature of the line 
emitting gas (NKK, Ballantyne et al. 2001).

(4) The abundance array for important elements, $A_Z$.

(5) The inclination of the disk with respect to the observer, $i$.

(6) The high energy spectral shape. It is commonly argued that 
this part of the spectrum does not affect much   the \fe line 
emission because very few photons in this energy range are 
responsible for K-shell photoionization. However, under the 
conditions of hydrostatic equilibrium considered herein, this 
part of the spectrum is instrumental in the determination of the 
Compton temperature of the ionized gas and therefore its ionization
state, thus impacting significantly the \fe emission.
See the Appendix where the role of the cut-off energy $E_c$ is
enunciated through several numerical tests.

Summarizing the foregoing discussion, the line flux is a function of:
\begin{equation}
\fline = \fline\left [ \fx, \Gamma, E_c, (\ldisk, A_Z, i)
\right]\;,
\label{fline}
\end{equation}
where we divided the variables into two groups. The second group of
variables (those in the curly brackets) is not expected to vary on
short time scales and hence these variables can be considered fixed in
\fe line reverberation studies. Indeed, $\ldisk$, the disk thermal
luminosity, is clearly the fastest varying parameter out of the second
group. It is expected to vary on the disk thermal time scale, which is
a factor $\alpha^{-1}$ longer than the disk dynamical time scale. So,
unless $\alpha$ is close to unity, $\ldisk\simeq$ const on the light
crossing time scales of the inner disk. On the contrary, we expect the
three parameters from the first group to be variable on roughly same
time scales on which $\fx$ varies. For example, when either
the geometry of the X-ray source or its luminosity varies, the
heating/cooling balance of the hot gas ought to change as well,
affecting values of $\Gamma$ and $E_c$.

Equation (\ref{fline}) clearly displays the challenges one would face
when analyzing \fe line reverberation data: In the case of more than
one parameter variations in the ionizing flux properties, the line
flux is not a function of any one single parameter but depends on all
of them. Therefore, studies of the correlation of the line flux with
the continuum flux {\em alone} may be insufficient to uniquely
determine the physical parameters of the problem and yield model
independent conclusions. We will come back to this point in \S
\ref{sect:rel}.

The high energy rollover is the parameter most difficult to measure on
short time scales. To circumvent this problem, one can concentrate on
targets that are relatively dim in terms of ratio $\lx/\ledd$ (but not
in terms of the observed X-ray flux!), since then no strongly ionized
skin is expected. Another approach is to pick AGN with $\Gamma >
2$. Although we have not presented any tests for these cases, it is
obvious that for steep X-ray spectra, the Compton temperature becomes
insensitive to the exact value and shape of the high energy roll-over
when $\Gamma-2$ is sufficiently large. Even if $E_c$ is highly
variable, the temperature and the Thomson depth of the skin will be
largely independent of its value.

\section{Time-dependent photo-ionized reflection: time scales}
\label{sect:tscales}

The problem of static X-ray illumination of an accretion disk involves
at least four major physical processes, none of which can be neglected
if an accurate solution to the problem is sought. These processes are:
(1) the ionization balance for the illuminated gas; (2) the balance of
heating and cooling; (3) the hydrostatic balance; and (4) the transfer
of the radiation through the slab. If the illuminating radiation field 

were constant for a long period of time and then (at time $t=0$ for
convenience) changed to a new intensity (spectral shape, etc) level,
then each of these four processes will be taken out of equilibrium. It
is thus necessary to ask how long it will take to re-establish the
equilibrium corresponding to the new ionizing intensity. This time
should then be compared to the light travel time across the inner
most disk region to see whether the non-equilibrium effects can be
important for \fe line reverberation studies.

The radiation transfer time scale, $t_{rt}$, is approximately
$(1+\tau_{i})\lambda_i/c$, where $\tau_i$ is the Thomson depth of the
ionized layer of material where most of the reflected spectral
features are formed, and $\lambda_i$ is the spatial extent of this
slab. The value of $\tau_i$ is a few (see, e.g., Ross \& Fabian 1993;
NKK), so $t_{rt}$ is only a factor of few longer than the light
crossing time $\lambda_i/c$.  The remaining three time scales were
discussed by Krolik, McKee \& Tarter (1981; \S IIa) for AGN line
emitting clouds, and we will simply re-scale their estimates for the
ionized skin. These authors found the radiative recombination time
scale for a hydrogenic ion of charge $Z$ to be about $1.6\times 10^8
T_{7}^{3/2}/Z^2 p_{13}$ sec, where $p_{13} \equiv nT/(10^{13} {\rm
cm}^{-3} {\rm K})$ and $T_7\equiv T/10^7$ K. This time scale is 
longest in the ionized skin (rather than in the cold dense material
below it), because there the gas temperature reaches its maximum while
the pressure has its minimum. We can estimate $p_{13}$ from the fact
that the ionization parameter in the skin is very high, i.e., $\xi
\equiv 4\pi \fx/n_H \simgt 10^{4}$ (e.g., Ross, Fabian \& Young 1999;
$\fx$ is the ionizing X-ray flux, and $n_H$ is hydrogen
density). Using $\fx \simeq L_{\rm x}/4\pi R^2$ where $L_{\rm x}$ 
is the total disk (or lamppost) X-ray luminosity, and $R$ is the 
radius under consideration, we obtain
\begin{equation}
\trec \simeq 0.2 \, T_7^{1/2}\, M_8^2\, r^2\, \xi_4\, Z^{-2}\,
L_{46}^{-1}\,{\rm s}\;,
\label{tr}
\end{equation}
where $M_8$ is the black hole mass in units of $10^8$ solar masses,
$r$ is the radius in units of \sch radius, $R_S$, $\xi_4\equiv
\xi/10^4$, and $L_{46}\equiv L_x/10^{46}$ erg$^1$ s$^{-1}$. To put
this estimate into perspective, one should compare it to the light 
crossing time, $R/c$:
\begin{equation}
R/c = 10^3\, r\, M_8\; {\rm s}\,.
\end{equation}

The hydrostatic time scale, $\thydro$, is $\lambda_s/c_s$ where
$\lambda_s \simeq [4 kT R^3 /GMm_p]^{1/2}$ and $c_s$ is the sound
speed, and hence
\begin{equation}
\thydro = 2^{3/2}\, r^{1/2} R/c\;.
\label{td}
\end{equation}
It is worth noticing that this time scale is always longer than
the light crossing time.

Finally, the thermal time scale can be shown to be longest on the
top of the ionized skin, where heating and cooling are dominated by 
the
Compton scattering. Writing $\tth = 3 m_e c^2/ U_{\rm rad} c 
\sigma_T$,
where $\sigma_T$ is Thomson cross section, and $U_{\rm rad} \sim
\fx/c$ is the radiation energy density, we have
\begin{equation}
\tth \simeq \frac{3}{2} r \frac{m_e}{m_p} \dm_x^{-1} R/c\;,
\label{tth}
\end{equation}
where $\dm_x = L_x/\ledd$. Note that if $r \sim 10$ and $\dm_x\simlt
10^{-2}$, then this time scale can be longer than the light crossing
time scale. Fortunately, for such low values of $\dm_x$, the Thomson
depth of the skin is small (see, e.g., Nayakshin 2000a) and thus the
existence of the skin is not important: the reflector is effectively
cold. Thus, in practical terms, it is reasonable to fix our attention
on the cases when $\tth\ll R/c$.

Let us now review these simple estimates in the light of efforts to
measure \fe line reverberation in AGN. Unless the observer is exactly
pole-on and the X-ray luminosity is a Dirac $\delta$-function in
time, one always samples emission from a range of
radii. Variability of the signal over time scales less than $R/c$ will
be difficult to measure, so that it only makes sense to talk about
variability on time scales longer than the light crossing
time\footnote{the characteristic value of $R$ is the radius within
which a significant fraction of the X-ray intensity is reprocessed by
the disk. For example, in the lamppost model, with height of the
source $h_x$ above the black hole, $R\sim h_x$}.
According to the foregoing discussion, we have
\begin{equation}
\pmatrix{ t_{\rm rt} \cr \trec \cr \tth\cr } \ll \frac{R}{c}\ll 
\thydro\;.
\end{equation}
The general problem of the time-dependent X-ray
illumination of accretion disks can be solved assuming that the
radiation field, ionization and thermal balances adjust to the new
value of the ionizing flux instantaneously\footnote{i.e., one can use
steady-state assumption for these processes}, but the gas density is
to be found from gas dynamical equations rather than from the
hydrostatic balance condition.

\section{Tests}\label{sect:tests}

\subsection{Setup}\label{sect:setup}

In the previous section, we argued that the time-dependent X-ray
illumination of accretion disks needs to be studied with a code that
performs radiation transfer, ionization and energy balance
calculations and also computes the dynamics of the gas.  We are not
aware of the existence of a code that would include treatments of all
of these processes to the level of details needed for \fe line
reverberation problem. However, we can use the static X-ray
illumination code described by Nayakshin, Kazanas \& Kallman (2000),
which is appropriate for equilibrium situations, to make certain tests
that can give us a rough idea of when time-dependence is important and
what kind of spectral changes are to be expected.

The tests that we will perform are of two types. In the first one, we
study situations in which X-ray flux was at one (static) luminosity
state which we will label (i) for time $t < 0$, and then switched
abruptly to a new state (j) at $t=0$ and remained at this state
indefinitely (i,j here are either 1 or 2).  Using the static NKK code,
we can compute the reflected spectra and the gas density profiles for
the states (i) and (j). In addition, according to \S
\ref{sect:tscales}, at time $\ttherm\ll t\ll \thydro$ we can compute
the reflected spectra under the assumption that the gas density
profile is still the same as it was in the state (i), but with the new
value for the illuminating flux. We will call this state (ij) as it
corresponds to the transition from state (i) to (j). It is clearly
important to examine whether reflected spectrum (ij) differs from that 
of
states (i) and (j).

The second type of tests that we will perform are less rigorous in the
mathematical sense but are not less interesting due to their
implications. In particular, we shall build a toy model for the time
evolution of the gas density in the illuminated slab when the X-ray
flux is variable.

Here and later in the paper, we choose the geometry of the lamppost
model, in which the X-ray source with luminosity $L_x$ is on the
symmetry axis above the plane of the accretion disk at height $h_x = 6
R_S$. The dimensionless accretion rate through the Shakura-Sunyaev
disk is $\dm = 0.03$. The illuminating X-ray spectrum is a power-law
with photon spectral index $\Gamma = 1.8$ extending to $E_c = 200$
keV. We further assume that the state 1 produces X-ray luminosity
equal to $L_{x1} = 0.3 L_d$, where $L_d$ is the total thermal
luminosity of the accretion disk. The state 2 is assumed to have
$L_{x2}=L_{x1}/2$.

\subsection{Reflected spectra immediately after the transition}
\label{sect:immed}

\begin{figure*}[H]
\centerline{\psfig{file=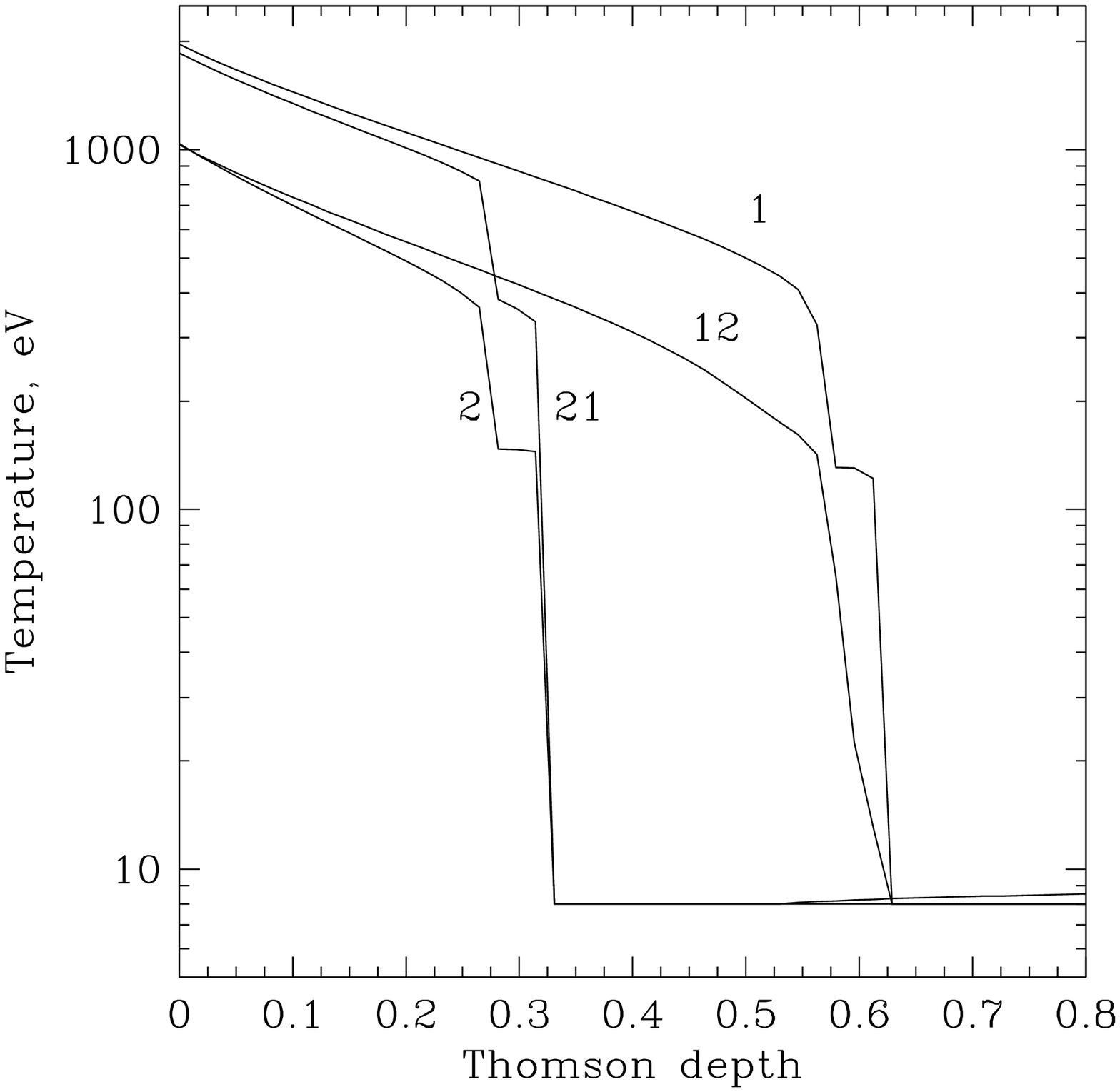,width=.9\textwidth,angle=0}}
\caption{Temperature profiles for the illuminated layer of the gas:
(1) and (2) are computed assuming the hydrostatic balance and the
``nominal'' values of the X-ray flux, $\fx$, (1) and (2), respectively
(see text); (12) is computed assuming the same density profile as in
(1) but with value of the illuminating flux appropriate for the state
(2). Similarly (21) is the same as (12) but for the transition from
state (2) to (1). Note that the differences in the temperature
profiles are highly significant observationally (see the following
figures).}
\label{fig:temp}
\end{figure*}

{\bf Flux increase.\ }
Figure \ref{fig:temp} shows temperature profiles resulting from the
calculation in which the initial state is (2) and the final state is
(1). The state (21) is defined to be the transient state which exists
for $t\ll \thydro$ after X-ray flux changed from the state (2) to (1).
Immediately after the transition, the Thomson thickness of the skin
remains the same, but its temperature increases roughly by a factor of
two, quickly adjusting to changes in the radiation
field.\footnote{Note that the temperature of the cold layer below the
skin does not change in the figure, but this is due to a convention
used by us to treat the low temperature solution. Namely, XSTAR
version 1 employed by us does not treat correctly certain many-body
physical processes important when LTE is expected (the most recent
version of XSTAR resolves this problem), which may lead to an
over-estimate of cooling at low temperatures. Therefore, following NKK
and \zycki et al. (1994), we set the temperature of the gas to be $kT
= 8$ eV if XSTAR returns temperature below this value. Since this is a
rather low temperature, the X-ray part of the spectrum is not expected
to be affected by this approximation. In reality, one anticipates the
temperature of the cold solution to vary by roughly a factor of
$(L_{x1}/L_{x2})^{1/4}\simeq 1.2$} This increase is easily understood
from the fact that the Compton temperature is proportional to $\fx$:
$T_c \propto J_x/J_{\rm disk}\propto \fx$, when $\fx\ll\fdisk$ (e.g.,
Nayakshin \& Kallman 2001). 

\begin{figure*}[t]
\centerline{\psfig{file=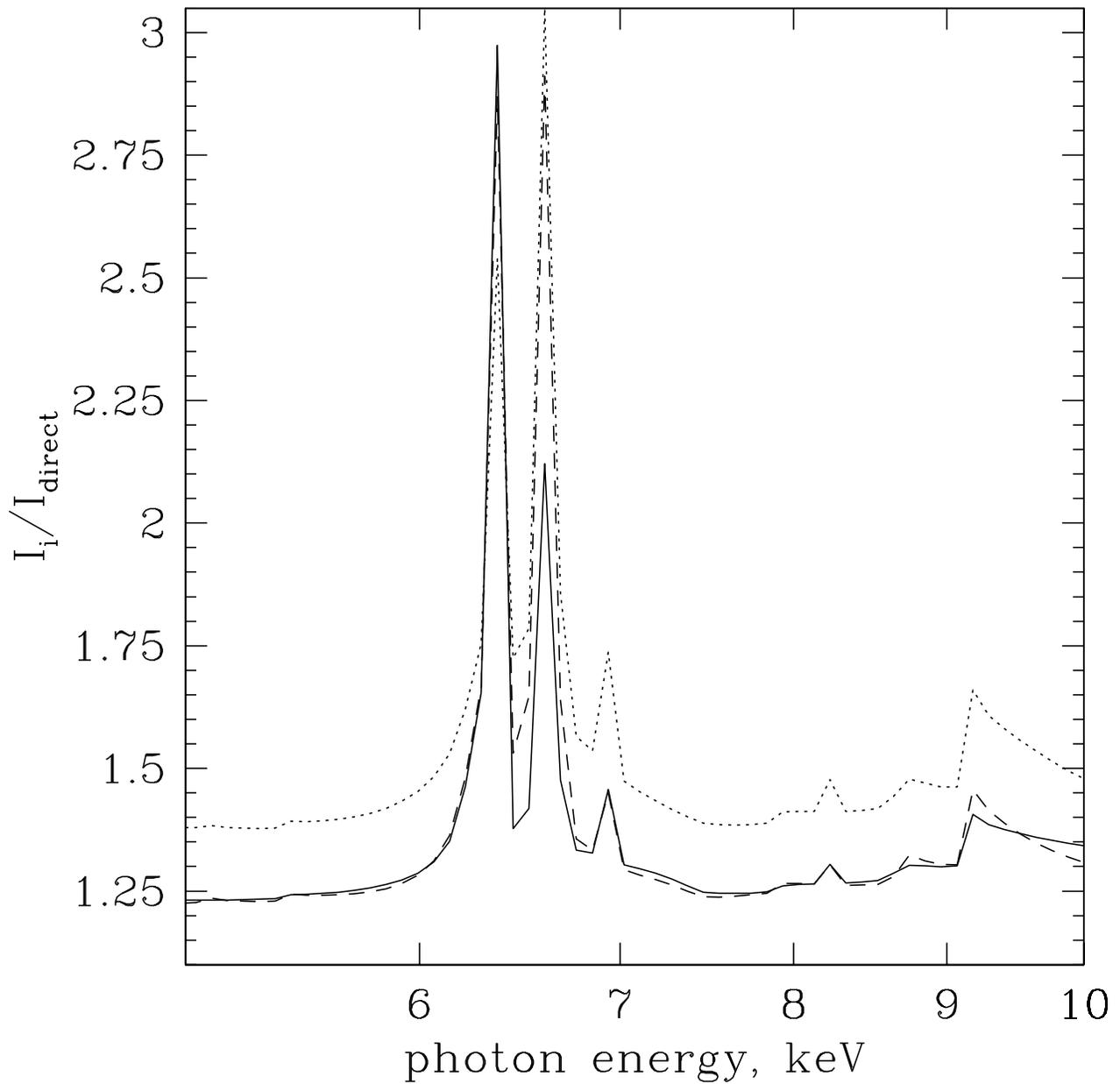,width=.9\textwidth,angle=0}}
\caption{Reflected face-on spectra for the hydrostatic balance models
-- (1) and (2), dotted and dashed, respectively, and the transient
spectrum (21), shown with the solid line. Note that \fe line complex
for the latter is distinctly different from that for either state (1)
or (2).}
\label{fig:spectr21}
\end{figure*}

Figure \ref{fig:spectr21} shows equilibrium spectra (1) and (2) and
also the reflected spectrum of the transient state (21), all for the
nearly-face-on viewing angle around Fe recombination band.  Note the
decrease in the He-like component of the line in the spectrum (21)
compared with that of state (2). This is due to higher temperature of
the skin which leads to correspondingly higher degree of
ionization. The integrated equivalent width of the \fe line complex
decreases from about 230 eV to 160 eV -- a change of roughly 30\%.

\begin{figure*}[t]
\centerline{\psfig{file=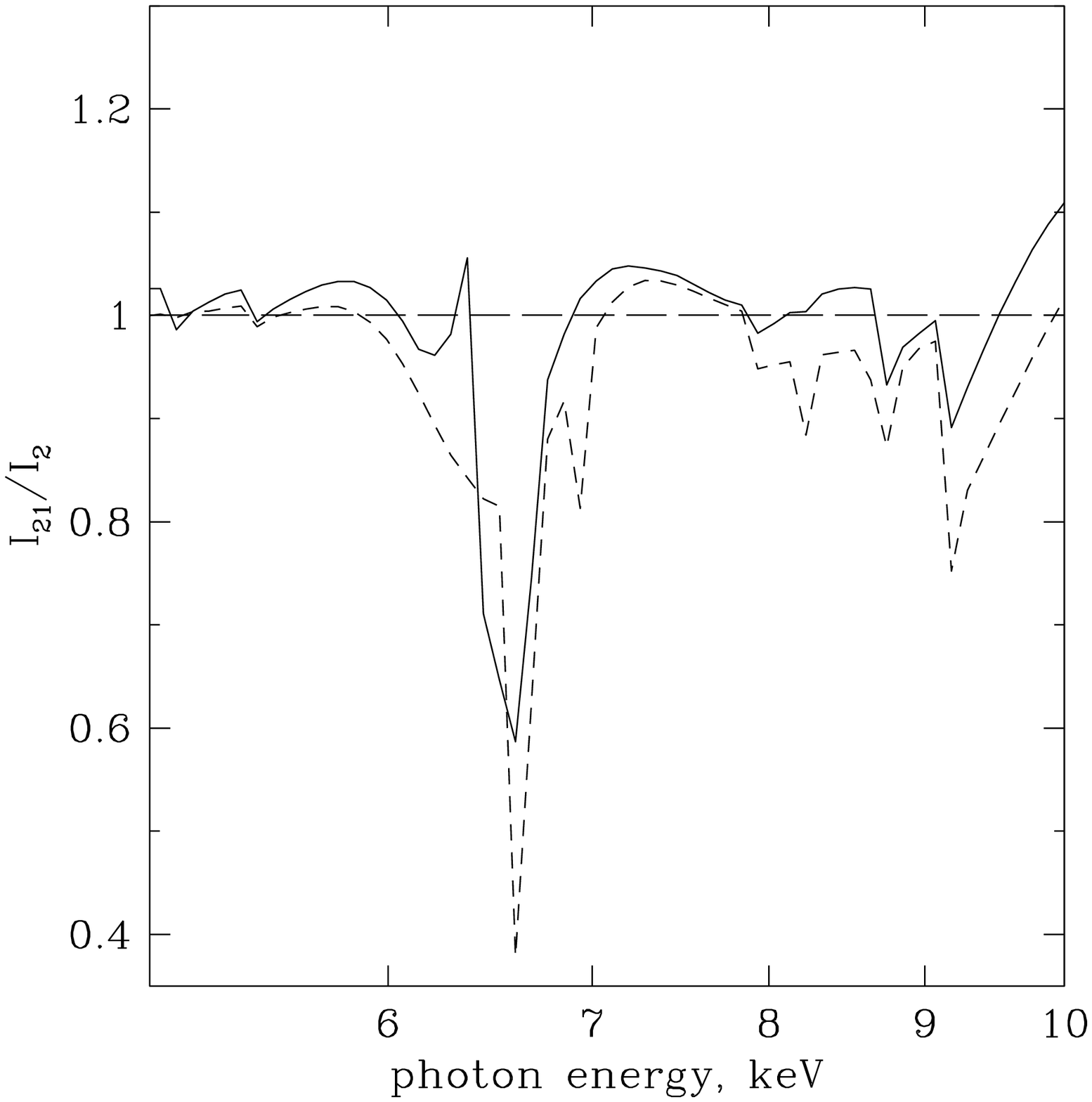,width=.9\textwidth,angle=0}}
\caption{Ratio of the transient reflected spectrum $I_{21}$ to that of
the initial state $I_2$ for two inclination angles: $\mu = \cos i =
0.95$ (face-on; solid curve) and $0.05$ (nearly edge-on; short dashed
curve). Note the deficit of the line emission from the medium to
highly ionized species of iron for the face-on spectrum and the
corresponding decrease in the highly ionized components as well as
their Compton-scattered wings for the edge-on spectrum.}
\label{fig:ratio21}
\end{figure*}

A slightly different representation of the spectral shape of the
reflected radiation is given in Figure \ref{fig:ratio21}, where we
plot the ratio of the reflected transient spectrum (21) to that of the
equilibrium spectrum (2) at two angles, one nearly face-on and the
other nearly edge-on. As already noted, the transient spectrum shows a
deficit of the intermediate to highly ionized He-like lines when 
viewed
face-on. Viewed edge-on, it displays a lack of He- and H-like lines
compared with equilibrium spectrum (2).  Note also a reduction in the
broad component of the line.  This component is due to Comptonization
of the line in the skin and is reduced simply because there is less
line emitted (and not because Comptonization becomes weaker).
 
{\bf Flux decrease.\ } Let us now consider the opposite case --
transition (12) from the ``high'' luminosity state (1) to the ``low''
luminosity state (2).  The corresponding temperature profile is also
shown in Figure \ref{fig:temp}. It is notable that in the ``transient
region'' -- the region with the Thomson depth $0.35 \simlt \tau_T
\simlt 0.6$ -- the state (12) actually occupies the solutions that are
thermally unstable, -- i.e., they are positioned below the point (c)
on the corresponding S-curve (see Fig. 1 in NKK). These solutions are
allowed, however, during the transition because the pressure balance
is not obeyed and the gas density is basically fixed (on short time
scales). At a fixed density, the thermal ionization instability does
not operate (see Krolik et al. 1981 and Field 1965).

\begin{figure*}[t]
\centerline{\psfig{file=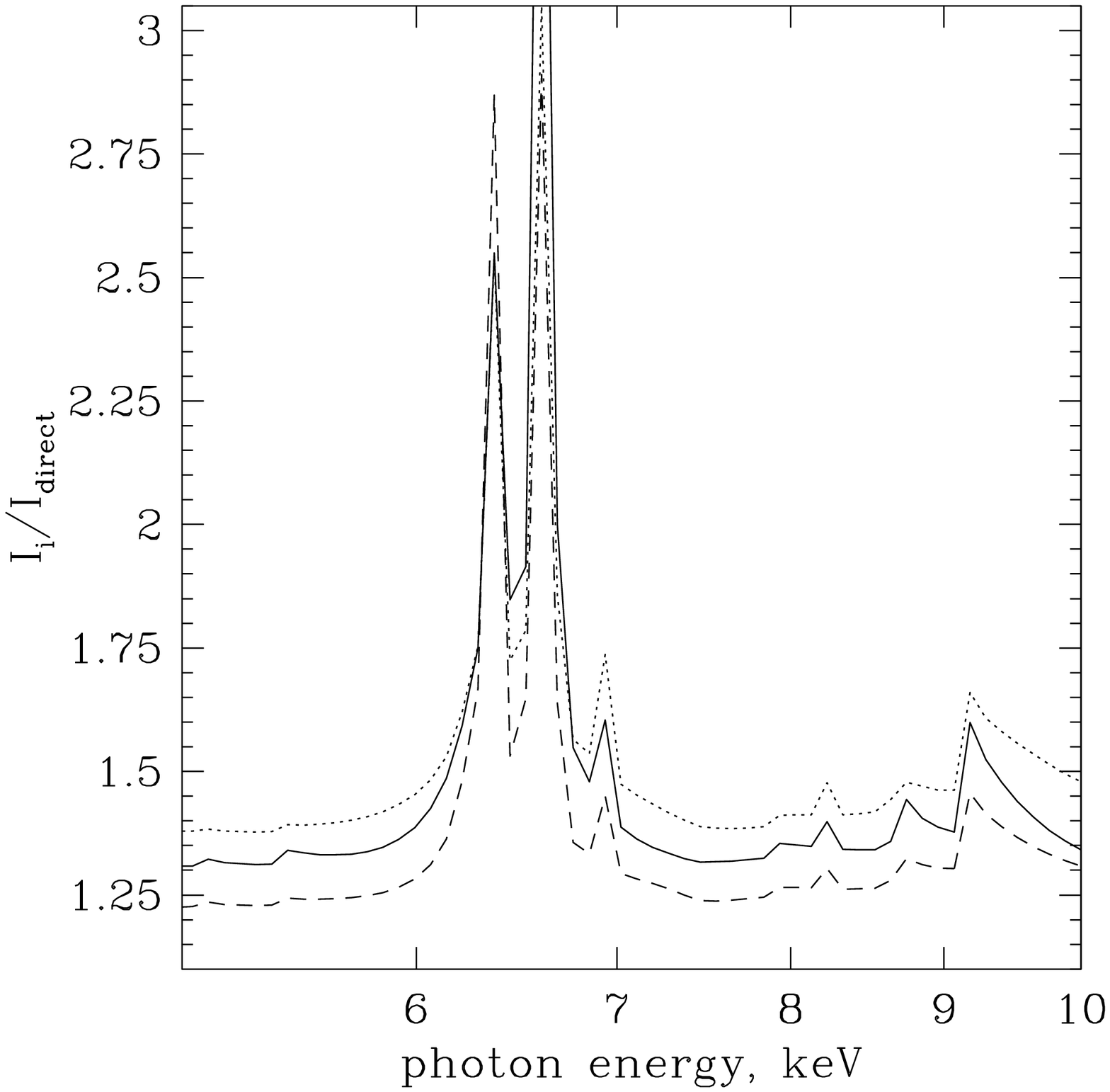,width=.9\textwidth,angle=0}}
\caption{Same as Fig. \ref{fig:spectr21} but for the transient state
(12) that is shown with the solid curve.}
\label{fig:spectr12}
\end{figure*}

\begin{figure*}[t]
\centerline{\psfig{file=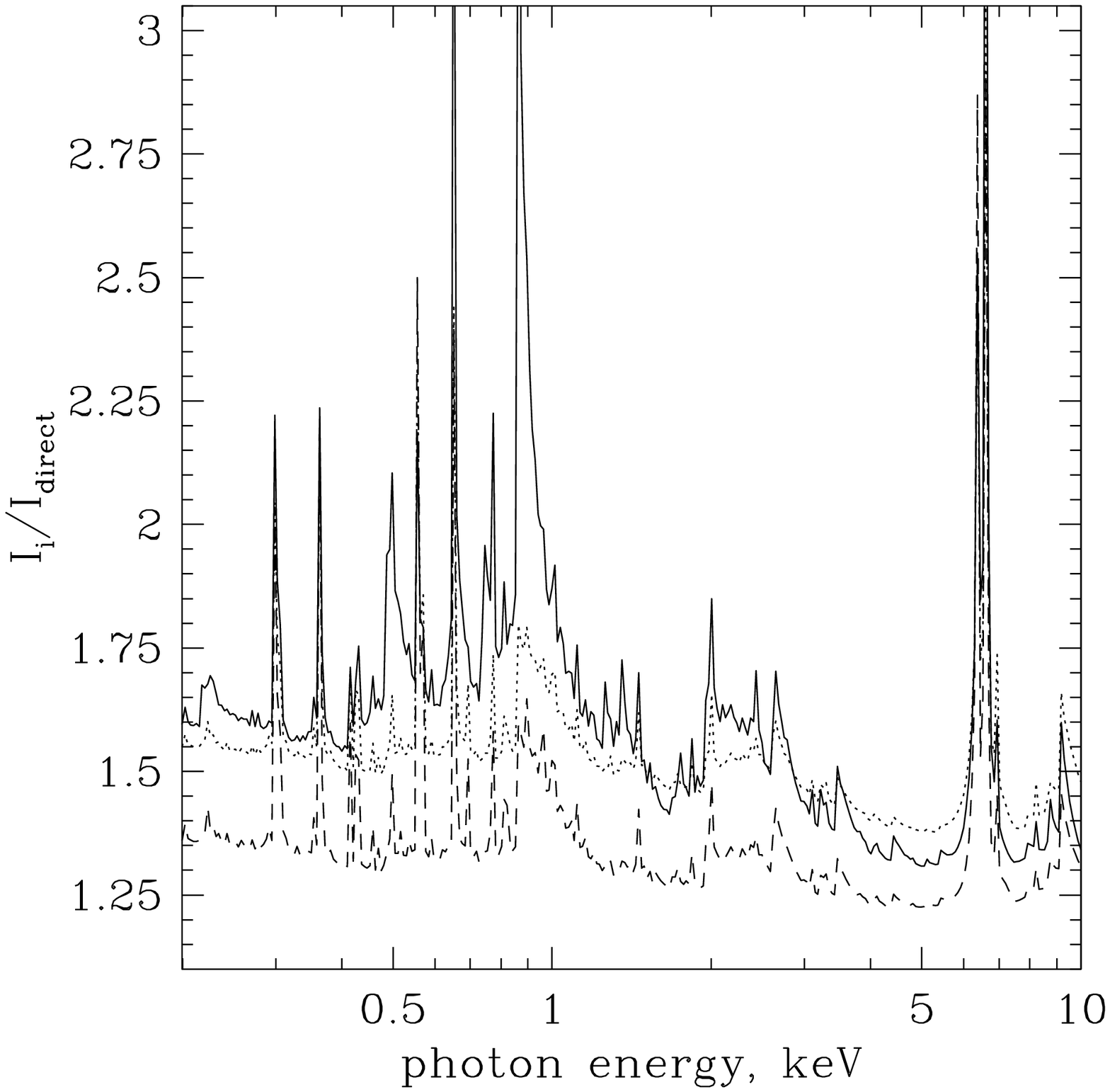,width=.9\textwidth,angle=0}}
\caption{Same as Fig. \ref{fig:spectr12} but in a broader energy
range.}
\label{fig:spectr12b}
\end{figure*}

Figure (\ref{fig:spectr12}) displays the transient spectrum (12) along
with the equilibrium spectra for comparison, while Figure
(\ref{fig:spectr12b}) shows the same spectra but in a broader
frequency range.  Clearly, the transient region yields significant
column depths in species from FeXVII to the completely ionized
FeXXVII.  The line emission from ``intermediate'' ionization stages,
FeXVII to FeXXIII, and the Compton down-scattered He-like line, fills
in (somewhat) the gap between the neutral-like component of the line
(that comes from the material beneath the ionized skin) and the
He-like component.

However, the most striking out-of-equilibrium effect is the increase
in the He-like component of \fe line. It is the change in the
emissivity of this feature that fuels the increase in the overall EW
of the line from 185 eV to 341 eV -- an increase of about
90\%. Clearly, an effect of such a large magnitude is observationally
significant.  Also, the changes in the soft X-ray band are even more
pronounced (Figure \ref{fig:spectr12b}). Variability in this energy
range is driven by variations in the ionization state of Carbon,
Nitrogen, Oxygen and Fe L lines. Both figures indicate that, as with
the transition (21), the continuum reflected spectrum (12) is also
different from either of the two equilibrium reflected spectra.

\subsection{A toy model for the density evolution of the ionized
skin}\label{sect:toy}

As we discussed above, in time-dependent situations, the reprocessing
gas density (at each point of the slab) is to be determined from
the associated gas dynamics, preferably using a gas-dynamical code. In
the absence of such a code, we will use the following simple
prescription:
\begin{equation}
\frac{\partial \rho(\tau_T, t) }{\partial t} =
- \frac{\rho(\tau_T) - \rho_{\rm eq}(\tau_T)}{\thydro}
\label{doft}
\end{equation}
where $\rho_{\rm eq}$ is the ``equilibrium'' state density profile of
the illuminated gas [for example, if we are studying the (12)
transition, then $\rho_{\rm eq}(\tau_T) = \rho_2(\tau_T)]$. Note that
$\rho_{\rm eq}(\tau_T)$ is a function of the Thomson depth (as opposed
to a constant), which ensures that the correct hydrostatic
balance-based solution is reached at $t = \infty$. While equation
(\ref{doft}) is admittedly ad-hoc, it mimics several important
properties of the density evolution that we expect.  Namely, at
times $t\ll \thydro$, the gas density profile is close to that of the
initial state; at times $t\gg \thydro$, $\rho(\tau_T) = \rho_{\rm
eq}(\tau_T)$; at intermediate times $t\sim \thydro$ the density
profile is a mixture of the initial and final states. In addition, the
re-adjustment of the gas density occurs on roughly correct time
scales. We plan to calculate the gas dynamics explicitly in our future
studies.

\subsubsection{Warm skin limit}\label{sect:warm}

Nayakshin \& Kallman (2001) have shown that the ionization state of
the skin and the reflected spectrum are strongly affected by the ratio
of the incident X-ray flux, $\fx$, to the disk flux, $\fdisk$. They
defined the ``warm skin'' limit in which the skin is highly but not
completely ionized. For hard X-ray spectra, $\Gamma \sim 1.8$ with
$E_c \simgt 100$ keV, this limit occurs when $\fx\simlt \fdisk$.  \fe
line emission in this case is dominated by Helium-like Fe at energy
$E\simeq 6.7$ keV (unless the skin is Thomson thin). The opposite
limit was termed the ``hot skin'' one: in this case the skin is nearly
completely ionized and it is possible to neglect all the atomic
processes in the skin. This limit holds for $\fx\gg \fdisk$ (for the
same spectra).

\begin{figure*}[t]
\centerline{\psfig{file=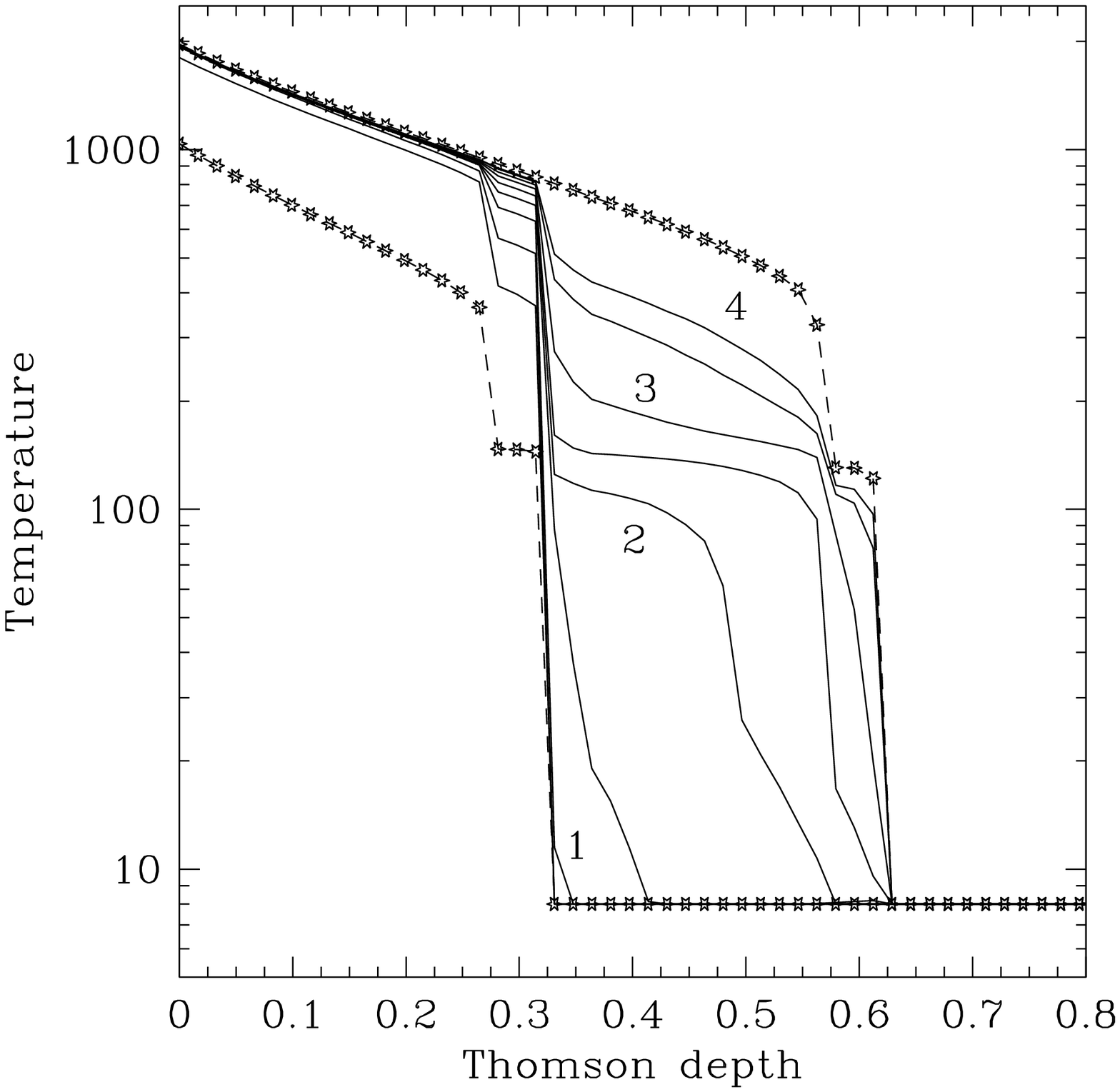,width=.9\textwidth,angle=0}}
\caption{Evolution of the temperature profiles for the illuminated
layer of the gas during the transition (21) computed under the
assumption that the gas density profile evolves as described by
equation (\ref{doft}).  The curves shown with the stars connected by
dashed lines are the equilibrium temperature profiles (1) and (2)
[same as in Figure \ref{fig:temp}]. Subsequent curves are computed for
times $t/\thydro = 0.5, 1.0, 1.5, ..., 4.0$. The numbers next to the
respective curves indicate time.}
\label{fig:temp21t}
\end{figure*}

Using this setup, we computed time evolution of the gas density and
the resulting spectra for transition (21). The calculation was carried
out with a time step of $0.1 \thydro$ up to time $t = 4 \thydro$. The
time-dependent temperature profiles are shown in Figure
(\ref{fig:temp21t}), with numbers labelling the times at which the
snapshots were taken.  Note that even at $t=4 \thydro$, the
temperature and the density profiles still have not reached the
``final'' equilibrium state.  In the ``transient'' region, i.e., at
$0.3\simlt \tau_T \simlt 0.6$, temperature evolves from the cold
branch through all the intermediate values to the hot branch. As
discussed in \S \ref{sect:immed}, the gas can temporarily occupy
thermally unstable solutions that are forbidden in the equilibrium
configurations.

\begin{figure*}[t]
\centerline{\psfig{file=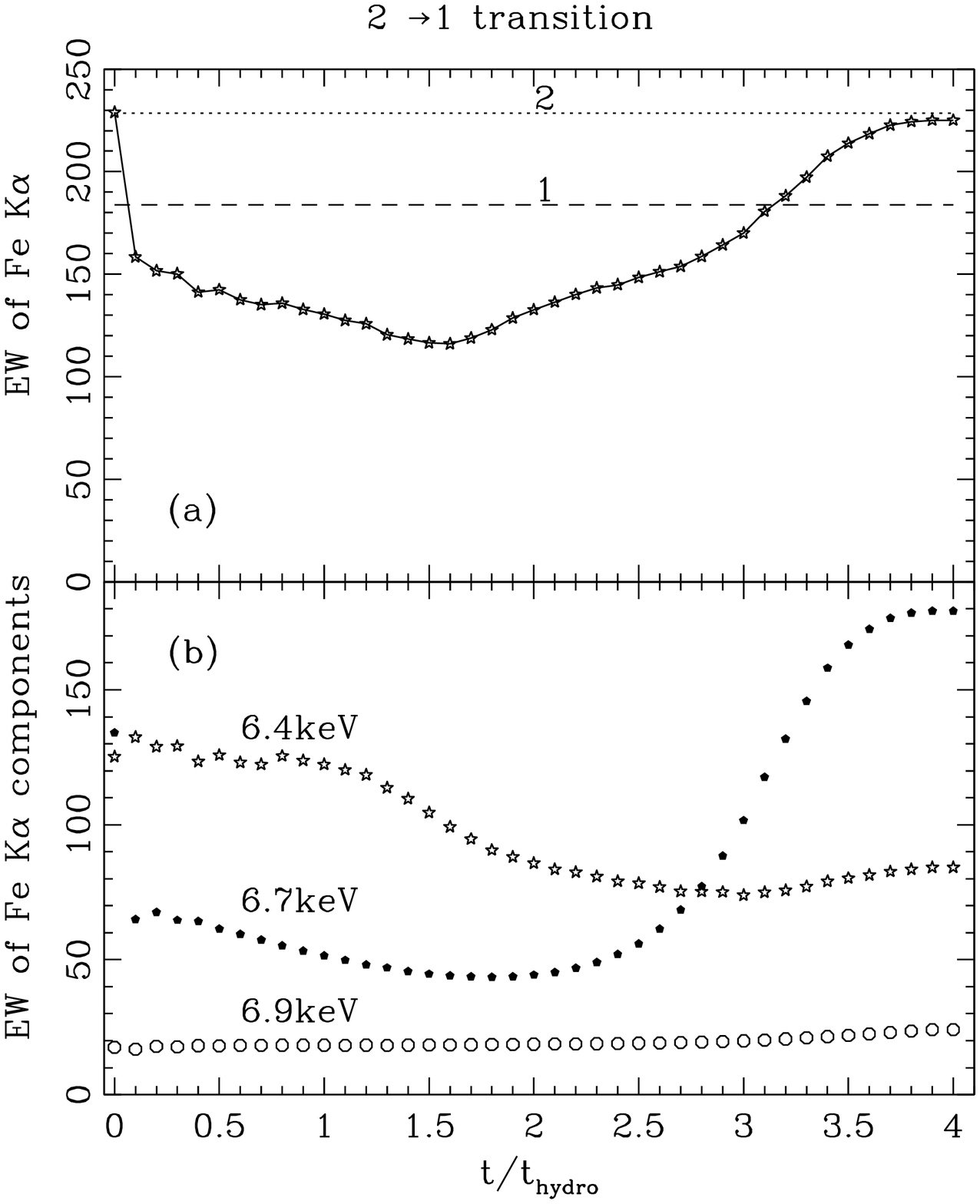,width=.8\textwidth,angle=0}}
\caption{(a) Stars -- Evolution of the total equivalent width (EW) of
the \fe line during the transition (21) as viewed at $\cos i =
0.95$. Horizontal lines show the EW of the initial and final
equilibrium states (1) and (2). Note that EW of the line still evolves
at the end of the simulation because the density profile has not
completely relaxed after $t= 4\thydro$, and that it will eventually
relax to the state (1). (b) ``Equivalent Width'' of
the bins that contain the neutral-like 6.4 keV component of the line,
Helium-like component at $\sim 6.7$ keV, and Hydrogen like one at
$\sim 6.9$ keV during the transition. The non-linear and non-trivial
character of the evolution of the EW with time is obvious.}
\label{fig:ew21t}
\end{figure*}

Panel (a) of Figure (\ref{fig:ew21t}) shows time dependency of the
equivalent width (EW) of the integrated \fe line profile, while panel
(b) displays the EW for the three prominent components of the line -- 
the
bins that contain the neutral-like 6.4 keV, the He-like $\sim 6.7$ and
H-like $\sim 6.9$ keV components. During the transition, the total EW
varies by a factor of 2, far greater than in the spectra obtained in
\S \ref{sect:warm} immediately after the change in the flux from (2)
to (1). This is clearly highly significant observationally since the
X-ray flux itself was varied by the same amount.

Let us now make a detailed analysis of the spectral evolution. From
Figure \ref{fig:temp21t}, we observe that at $t\simlt \thydro$, the
cold layers below the ionized skin are relatively unaffected. Though
in part this is due to the simplified treatment of the cold branch of
the solution (as described in the footnote to \S \ref{sect:immed}), we
believe this is a general result because the cooling of the cold
branch is done mostly through optically thick processes, and hence the
cooling function there depends very strongly and non-linearly on
temperature. A change in the X-ray flux of the order of 2 will cause
only a slight change in the temperature of this region, contrary to
the highly ionized skin (where the cooling function depends on the
temperature in an approximately linear fashion). In addition, the gas
temperature must increase to about $100$ eV, that is about 10 times
from its initial value of $8$ eV, in order for the iron to become
sufficiently ionized to permit FeXVII and higher ionization species to
become dominant line emitters. Therefore, the line emission from this
cold region -- the neutral-like component at $E = 6.4$ keV -- evolves
very slowly in the beginning of the calculation, as can be seen in
panel (b) of Figure (\ref{fig:ew21t}). Only at $t\simgt 1.5\, 
\thydro$,
the transient region's density evolves enough to allow for a
substantial warming up of that region and the associated decrease in
the neutral-like \fe line emission.

The evolution of the Helium-like \fe component is dictated by the fact 
that
at first the skin heats up due to the increase in the ionizing flux 
and
therefore the emission from this component significantly decreases. At
later times, when the transient region warms up, the He-like emission
increases by as much as a factor of 4. The Hydrogen-like component at
$E\simeq 6.9$ keV remains very weak and approximately constant during
the entire interval, a fact due mostly to the intrinsically low 
effective
fluorescence yield of H-like Fe. Also note that at times $t\sim 1-2
\thydro$, the soft X-ray band exibits many strong lines and
recombination continua which we do not show here
(cf. Fig. \ref{fig:spectr12b}; in addition, animations of the
temperature and spectral changes can be found at
``http://lheawww.gsfc.nasa.gov/users/serg/'').

\subsubsection{Hot skin limit}\label{sect:hot}

\begin{figure*}[t]
\centerline{\psfig{file=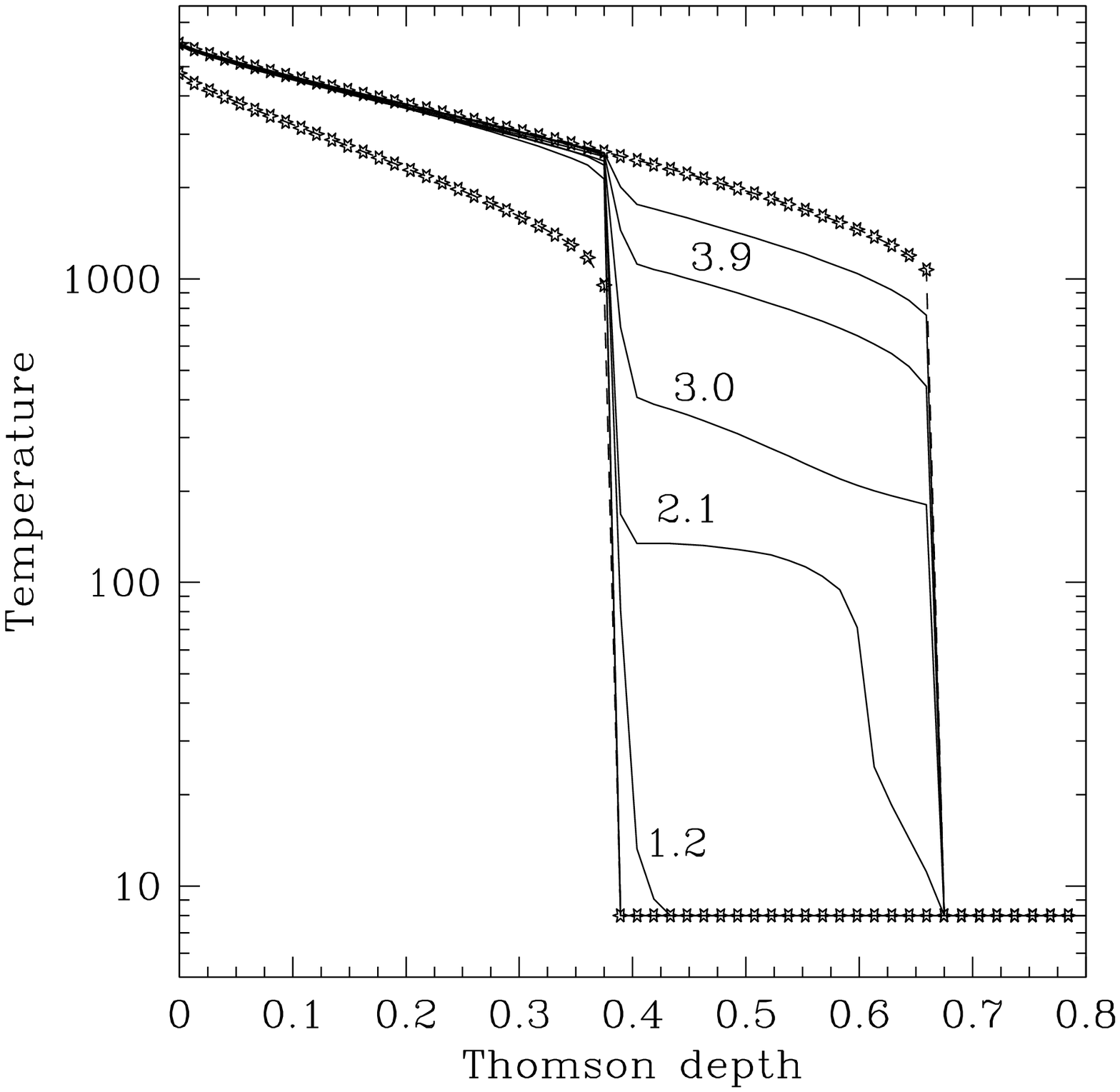,width=.9\textwidth,angle=0}}
\caption{Evolution of the temperature profiles for the illuminated
layer of the gas during the transition (21) in the hot skin limit (see
\S \ref{sect:hot}).  The meaning of the curves is the same as in
Figure \ref{fig:temp21t}]. Subsequent curves are computed for times
$t/\thydro = 0.5, 1.0, 1.5, ..., 4.0$. The numbers next to the
respective curves indicate time.}
\label{fig:temp21ht}
\end{figure*}

\begin{figure*}[t]
\centerline{\psfig{file=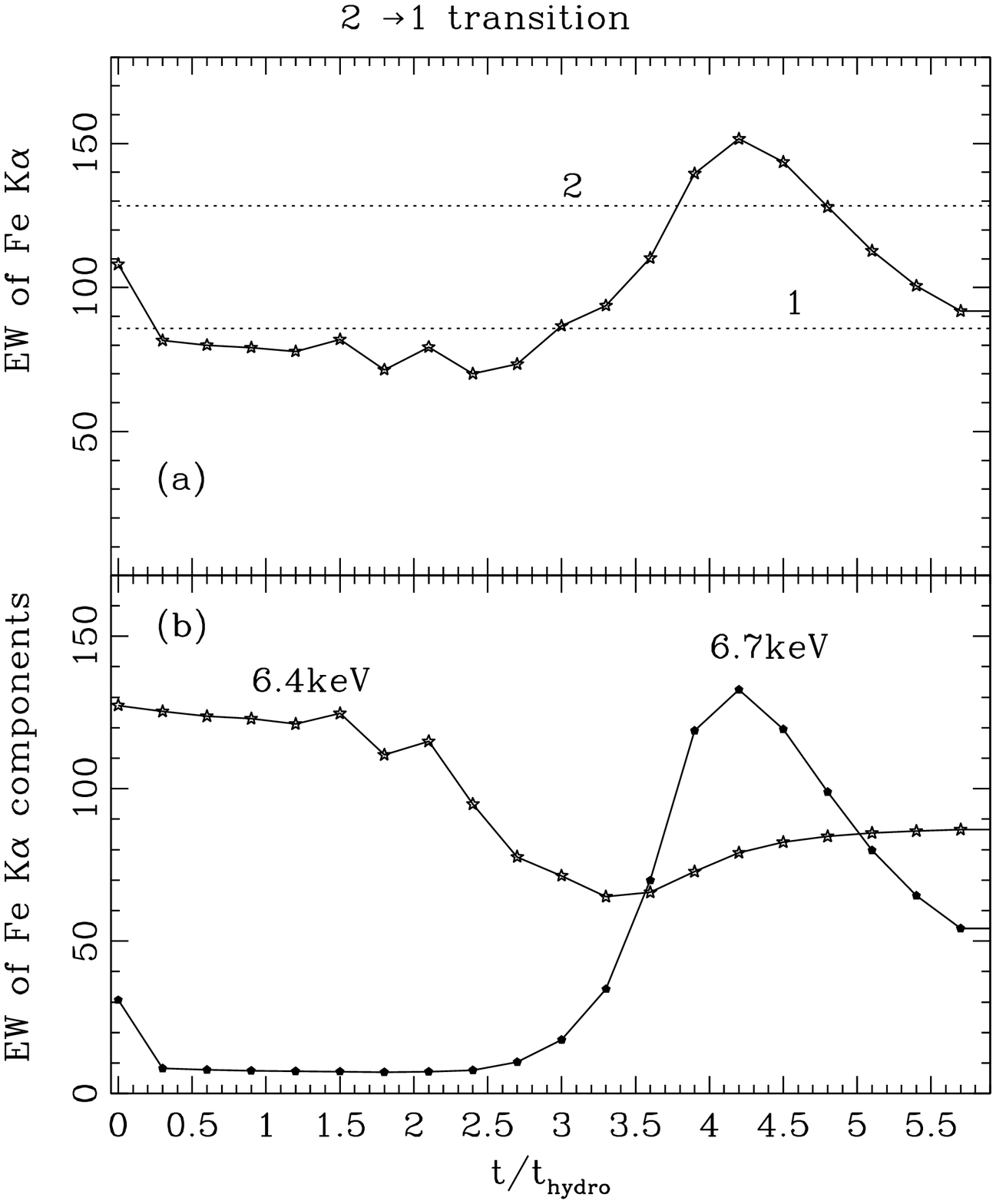,width=.9\textwidth,angle=0}}
\caption{Same as Figure \ref{fig:ew21t} but for the hot skin limit.
Hydrogen-like line emission at $\sim 6.9$ keV is not shown since it is
negligible at all times.}
\label{fig:ew21ht}
\end{figure*}

The second test we perform is for the same geometry, but for the ratio
of $\lx/\ldisk = 3$ in the state (1), and the accretion rate $\dm =
3\times 10^{-3}$. The corresponding temperature profiles are presented
in Figure \ref{fig:temp21ht}, while the EW of the line is plotted in
Fig. \ref{fig:ew21ht}. The character of the temperature and spectral
evolution during the transition is similar to that of the warm skin
limit, except that the He-like line emission suffers even greater
variations. This is a consequence of the fact that the skin in state
(1) is completely ionized, which means that the He-like line
emissivity evolves from virtually zero to being the dominant component
and then decaying as the skin approaches the new (hot) state (2).

\section{Discussion}\label{sect:disc}

\subsection{Main results and implications for current observations of 
AGN}
\label{sect:main}

Our main results can be summarized as following.

{\bf (1)} The gas hydrodynamical time, $\thydro$, is an important
    relaxation time scale of the X-ray reverberation problem. 
	The hydrostatic equilibrium assumption  fails for 
	variations with characteristic time scale $t \lesssim \thydro$.
	As shown, this will in general affect the reflected spectra, unless
    the Thomson depth of the ionized skin is much smaller than unity.

\begin{figure*}[t]
\centerline{\psfig{file=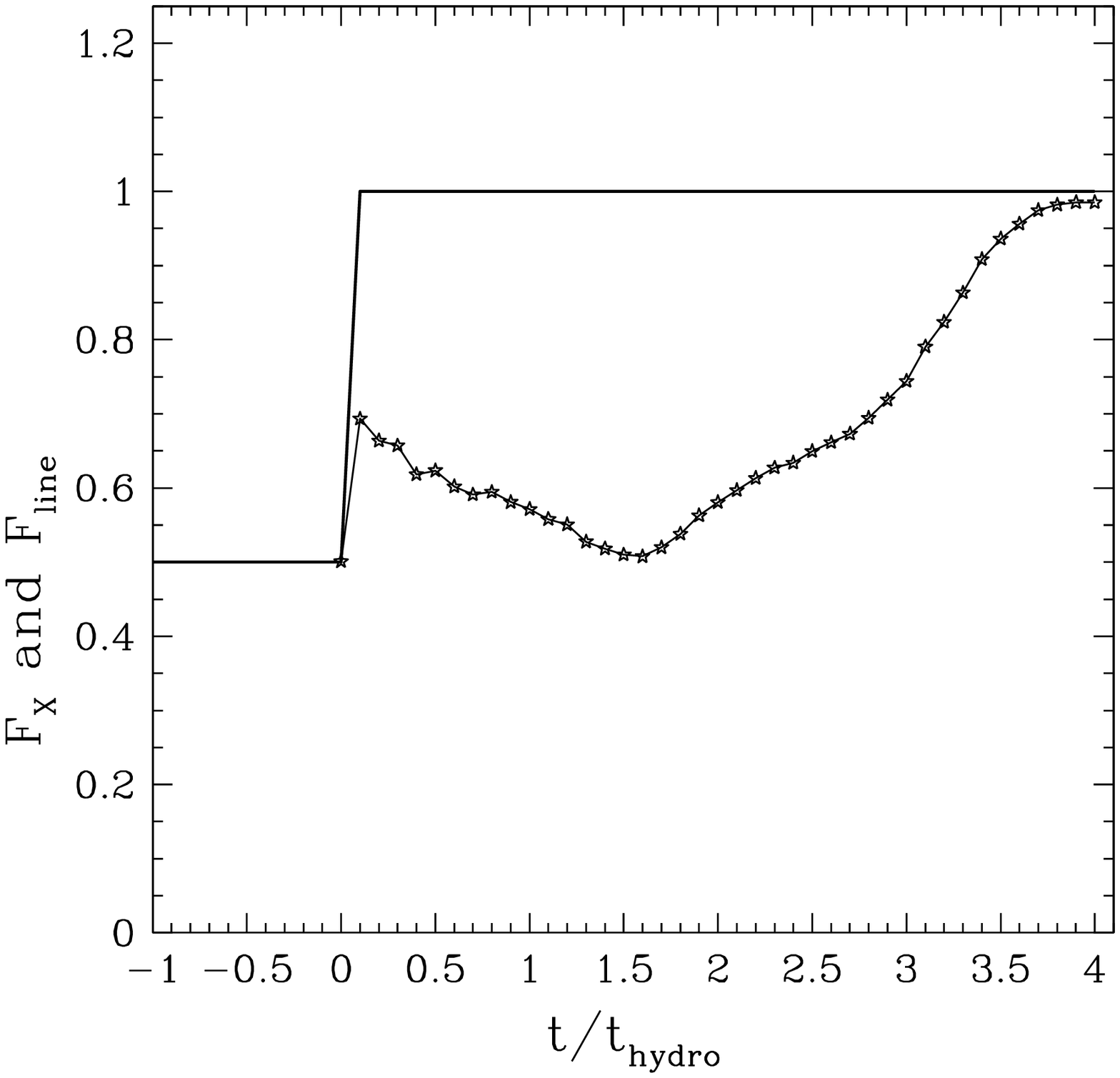,width=.9\textwidth,angle=0}}
\caption{The continuum X-ray flux and \fe line flux as a function of
time for transition studied in \S \ref{sect:warm}. Note that the line
flux goes up at substantially later time, and this ionization physics
effect may be misinterpreted as a geometrical delay.}
\label{fig:fline21t}
\end{figure*}

To make this point clearer, we plot (Figure \ref{fig:fline21t}) the
time profile of the illuminating flux (in slab geometry) and the 
line emissivity in normalized units for the simulation presented in \S
\ref{sect:warm}. It is apparent from this figure that the line flux
responded to the change in the illuminating flux only after a delay
$t\sim \thydro$. Therefore, there is a possibility that the slow
evolution of the \fe line emissivity, caused by the evolution of the
ionization state of the reflector, might be mis-interpreted as being
due to the specific geometric arrangement or the presence of
general-relativistic effects. We will discuss this point a little
further in \S \ref{sect:impl}.

{\bf (2)} If the illuminating flux fluctuates on time scales $t \gg
    \thydro$, the resulting spectra can be computed in the
    quasi-static approximation (e.g., NKK). Effectively, this is the
    limit in which all of the ionized X-ray reflection calculations
    (of which we are aware) were performed to date. 

{\bf (3)} In the opposite case, $t \ll \thydro$, the ionized disk is
	out of hydrostatic equilibrium and the reflected spectra can
	be significantly different from that obtained under the
	quasi-static assumption. In particular, the \fe line
	emissivity is {\em not} a function of the instanteneous X-ray
	flux that would have been implied from the hydrostatic models
	of NKK. The Thomson depth of the skin is approximately that
	appropriate for the time-averaged X-ray flux, but the
	transition from highly to weakly ionized layers is not sharp
	and time-steady as in the models of NKK. The time-averaged
	reflected spectra are {\em not} the same as quasi-static
	reflected spectra computed for the average incident flux.

\begin{figure*}[t]
\centerline{\psfig{file=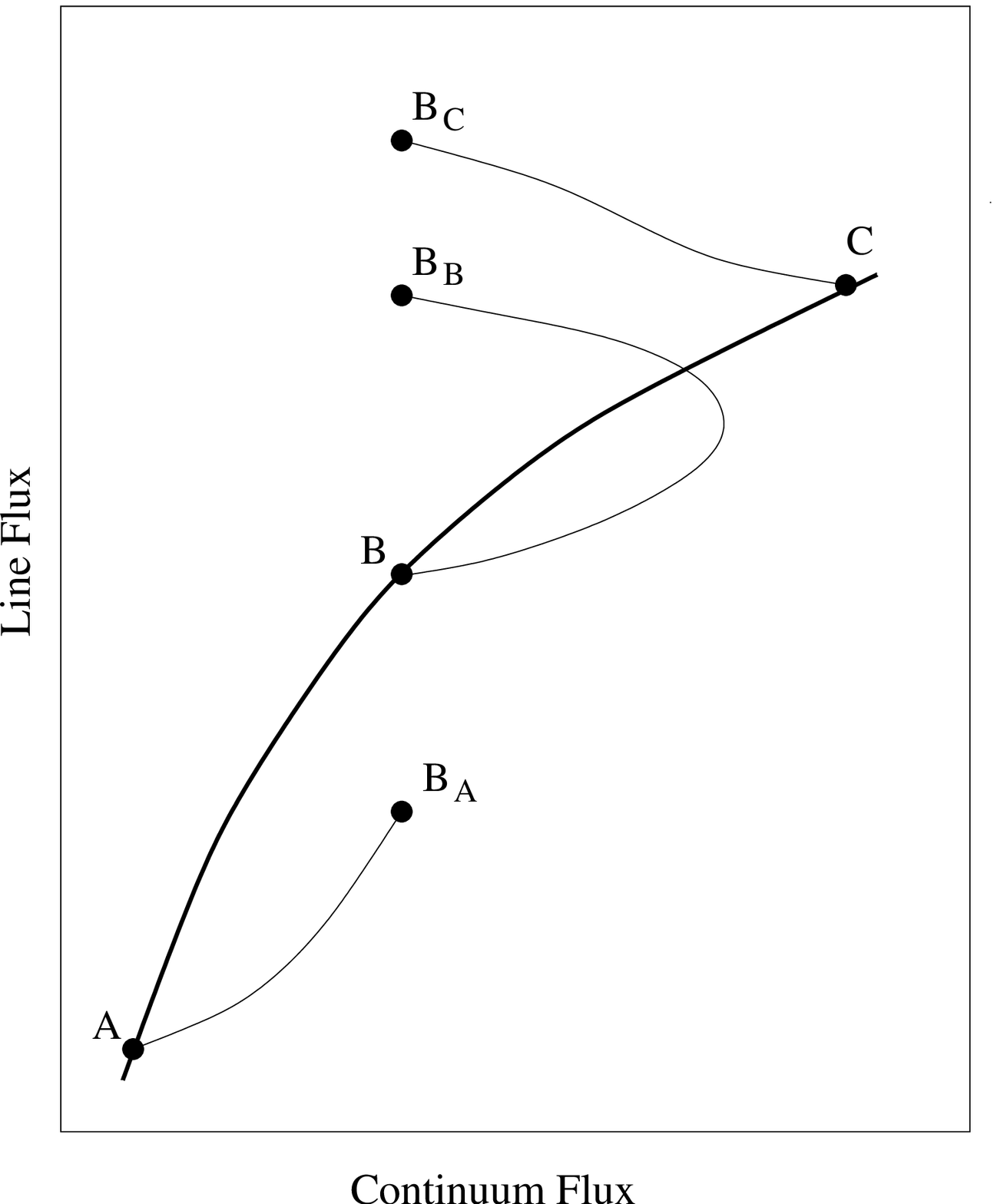,width=.5\textwidth,angle=0}}
\caption{Schematic drawing of the dependency of line flux on the
continuum flux. The thick solid curve shows the equilibrium curve --
i.e., one that is obtained for a fixed shape of the illuminating
spectrum and assuming hydrostatic balance. Non-equilibrium paths
A--B$_A$, C--B$_C$, and B--B$_B$ are shown as examples of the
``trajectories'' allowed under time-dependent conditions. Thus, in
non-equilibrium situations, more than one value of the line flux for a
given continuum flux is to be expected.}
\label{fig:draw1}
\end{figure*}

This is not surprising if one considers that the response of the disk
to the ionizing radiation is a strongly non-linear process driven by
the TII. The observed response depends not only on the frequency of
variation but also on the initial conditions of the reprocessing
matter. Consider, for example, a specific dependence between the
X-ray flux and the line equivalent width under steady-state conditions
depicted schematically by the thick solid line in Fig. \ref{fig:draw1}
(for realistic examples see Figure 1 of Nayakshin 2000b). For fast
flux variations, i.e., those for which changes in $\fx$ occur on time
scales shorter than $\thydro$, one will, in general, observe different
levels of \fe line flux if one arrives at the continuum flux
corresponding to point B starting from different initial points
(cf. paths A--B$_A$ and C--B$_C$ in the figure). In fact, since the
flux variations disturb hydrostatic balance, one will not come back to
the initial point on the curve even if the flux attain its original
value (see trajectory B--B$_B$). These curves will converge to the
same value of the ordinate only if the X-ray flux did not change from
its final value (at point B) for time scales much longer than
$\thydro$.  Hence, on short time scales there is no ``correct'' value
for the line flux at a given continuum level and therefore one could
obtain no apparent correlation between these two quantities in a
monitoring campaign, instead of the one expected on the basis of
steady state calculations.

This argument may provide an explanation 
for the results of the analysis of the \fe line variability of 
MCG 6--30--15 by Vaughan \& Edelson (2001): These authors split 
the X-ray light curve of this object into bins of the RXTE 
sampling period, i.e. of order of 6000 seconds. They found (see 
their Figure 3) that the \fe line flux was variable on time scales 
of a few RXTE orbits, but it was {\em not} correlated with the 
continuum flux in any consistent manner.

The parameters associated with this object are roughly consistent with
those needed for an explanation of these results as due to the finite
value of $\thydro$. Demanding the value of the latter to be $\simeq
6000$ sec and assuming a ``reasonable'' value of $r=5$, one obtains
(Eq. \ref{td}) for the mass: $M \sim 2 \times 10^7 \msun$.  The
observed X-ray luminosity of MCG 6--30--15 is $\sim {\rm a~few} \times
10^{43}$ erg s$^{-1}$ (e.g., Iwasawa et al. 1996), implying an
accretion rate of a few percent or more. Following the results of
Nayashin \& Kallman (2001) it is expected that the Thomson depth of
the ionized skin would be sufficiently large to make the
out-of-equilibrium effects significant.

Finally, one should bear in mind that the \fe line emissivity of an
{\em ionized} reflector is a function of the total radiation continuum
seen by the reflector and not just the $2-10$ keV part of it typically
picked by observers. In fact for spectra with $\Gamma < 2$ the $E
\simgt 10$ keV part of the spectrum contains most of the X-ray
luminosity and thus determines the depth of the ionized skin.  Thus,
\fe line flux and $F_{2-10}$ do {\em not} have to be directly related
even in the quasi-static case. Variations in this (unobserved) part of
the spectrum (see Eq. 1 and the Appendix) might of course be invoked
for a more prosaic explanation of the results of Vaughan \& Edelson
(2001).

\subsection{Implications for future Fe K$\alpha$ line
reverberation studies}
\label{sect:impl}

The question of immediate interest following the discussion above is
the potential impact of the non-equilibrium effects on the future \fe
line reverberation campaigns (e.g., see Young \& Reynolds 2000). The
Thomson depth of the skin is much smaller than unity in the lamppost
geometry if $\lx\ll 0.01 \ledd$, and then the ionized layer is not
important altogether. Many AGN are brighter than this, possibly, and
for these the non-equilibrium effects ought to be important. However,
the calculations presented above are purely local, considering the
response of only a small region of the disk in slab geometry. More
work is needed to determine whether the {\em local} effects will
affect the reflected signatures of a full disk. While this complicates
the theory of \fe line reverberation, we should not forget that with
that there comes a new tool (beyond static models) to distinguish
between models of different geometric arrangements on the basis of the
responce of the reflected spectra to the continuum.

Conceptually the simplest model is that with the lamppost geometry.
The X-ray source is stationary and it is the \fe line ``echo'' from
different parts of the disk that determines the reflected line
profile.  An alternative arrangement is that of magnetic flares
(Nayakshin \& Kazanas 2001). In this geometry the observed $\lx$ is
the aggregate of the emission of a large number of flares that take
part in Keplerian rotation of the disk. Because of the rather limited
extent of magnetic structures on the disk, one expects the \fe line
features to be rather narrow and oscillate in energy space with the
Keplerian frequency (Nayakshin \& Kazanas 2001; Ruszkowski 2000),
properties which may allow future observations of higher energy and
time resolution to distinguish them in the X-ray light curves.  As
long as the narrow features are distinguishable in the spectrum, their
magnitudes do not matter in the analysis of the trajectories of the
emitting spots, and this is why the complications with the
non-equilibrium effects may be circumvented.

\subsection{Frequency Resolved Spectroscopy}
\label{sect:gbhc} 

At this point one should note that the power density spectra of AGN as
well as those of GBHCs are power laws which cover a large number of
decades in Fourier frequency. In AGN specifically, they extend to very
low frequencies ($\lesssim \mu$Hz; Edelson \& Nandra 1999), much lower
than the characteristic time scales discussed in section
\ref{sect:tscales}; therefore, one would expect at some frequency the
transition from the $t < \thydro$ to the $t \gg \thydro$ regime, with
the concommitant change in the correlative properties between the
continuum and the \fe line.

Perhaps the most efficient way of studying these effects is to 
implement, when appropriate, the frequency resolved spectroscopy
method employed by Revnivtsev, Gilfanov \& Churazov (1999) in
the spectro-temporal analyis of the data of Cyg X-1 and compare
the results to models which incorporate the above discussed 
processes. In applying their method to the hard state of Cyg X-1, 
Revnivtsev et al (1999) found that the \fe line and reflection 
features appear to weaken with increasing in Fourier frequency. 
In particular, these features drop to about half of their value 
at a frequency $f\sim 10$ Hz, and continue to decline to the highest 
frequencies probed by their analysis ($f\sim 30$ Hz).

Our models predict that the one-to-one correlation between the X-ray
flux and the \fe line flux may be broken on short time scales. They do
{\em not} predict a decrease in the variability of \fe line with
increasing frequency as such, so the lamppost-like models do not seem
to offer any physical explanation for results of Revnivtsev et al
(2000).

At the same time, note that the observations also show that the X-ray
continuum slope decreases with increasing frequency from about
$\Gamma\sim 1.9$, for frequencies below $f\sim 1$ Hz, to $\Gamma\sim
1.6$ for $f\sim 30$ Hz. This is significant, because as shown by NKK,
Done \& Nayakshin (2001), Ballantyne et al. (2001), with other things
being equal, a steepening in the X-ray spectrum lowers the Thomson
depth of the skin and its degree of ionization. In particular, when
the X-ray spectrum is hard\footnote{the precise value of $\Gamma$
depends on $E_c$ as explained in the Appendix, but for Cyg~X-1 the
cutoff energy is quite high in the hard state, i.e., $E_c\sim 100-200$
keV}, i.e., $\Gamma \sim 1.7$, the skin is nearly completely ionized
so it supresses all the reflection signatures. For $\Gamma\simgt 2$,
the skin contains a non-negligible amount of He-like iron, and then
the line cannot be completely suppressed, no matter how high the
ionizing flux is.  These facts incorporated in the observations of
Revnivtsev et al.  (1999) might hence provide an account for the
corresponding behavior of the \fe line behavior observed within the
framework of the models of NKK, NK and Done \& Nayakshin
(2001). Clearly the regions processing hard flux should be
geometrically distinct from those processing softer flux, which is
possible only if these fluxes result from different X-ray sources
(e.g., magnetic flares).  However, a more definitive statement along
these lines will have to await the development of more detailed 
models.

\section{Conclusions}

We performed a first, highly simplified, calculation of time-dependent
spectra due to X-ray reprocessing on the surface of an accretion disk,
assuming a step function variation in the incident X-ray flux and
limiting ourselves to their effects on a limited section of the
accretion disk which we have treated in the slab approximation.  While
our treatment of the hydrodynamic response of the photoionized gas
density profile to changes in the X-ray flux is admittedly very
simple, we nonetheless expect that our {\em qualitative} conclusions
will hold in a more sophisticated calculation. We found that the time
necessary for establishing hydrostatic balance in the X-ray heated
skin of the disk is an important relaxation time scale of the problem.
Furhtermore, because of the non-linear nature of the optically thick
ionized X-ray reflection problem, the ``relaxation'' process allows
for entirely new physical solutions at transient times, absent in the
steady-state solutions. One of the new interesting features is the
presence of a strong He-like component at $\sim 6.7$ keV of the \fe
line during a transition from an initial neutral-like to the final
neutral-like spectrum.

Because of the presence of this relaxation time scale, under
time-dependent conditions, the reflected spectrum is {\em not} a
function of the instantaneous illuminating flux or other system
parameters and depends on the history of variations of these
parameters. This fact may provide an explanation for the recent
analysis of time-dependent reflection data for MCG-6-30-15 by Vaughan
and Edelson (2001).  It is also important to note that, because of the
non-linear character of the formation of the ionized skin on the
surface of the accretion disk, as indicated by our results, the {\em
time-integrated} spectra of real sources may not always be modelled as
static spectra corresponding to the average X-ray luminosity during
the observation.

While a definitive asnwer of the effects described above within
the specific models should await further more detailed investigations,
as argued in section \ref{sect:impl} they can be important for the
for the {\em lamppost models} of \fe line reverberation of
AGN once X-ray luminosity of the source exceeds a percent or so
of the Eddington luminosity. Their importance in other such models
e.g. the flare models of Ruszkowski (2000) and Nayakshin \& Kazanas
(2001) is harder to assess because of the additional details that
will have to be specified to completely determine these models.
Nonetheless, for any given model, these effects are well defined and 
their incorporation in the analysis of future observations of higher 
time and energy resolution should provide additional means for 
discriminating between the various models and hence contribute to 
probing the physics of accretion in the vicinity of compact objects.

SN acknowledges support from the National Research Council during the
bulk of the time devoted to this project. The authors are very
grateful to Tim Kallman, Chris Reynolds and Mitch Begelman for
discussions and useful comments that improved this paper.

\begin{figure*}[H]
\centerline{\psfig{file=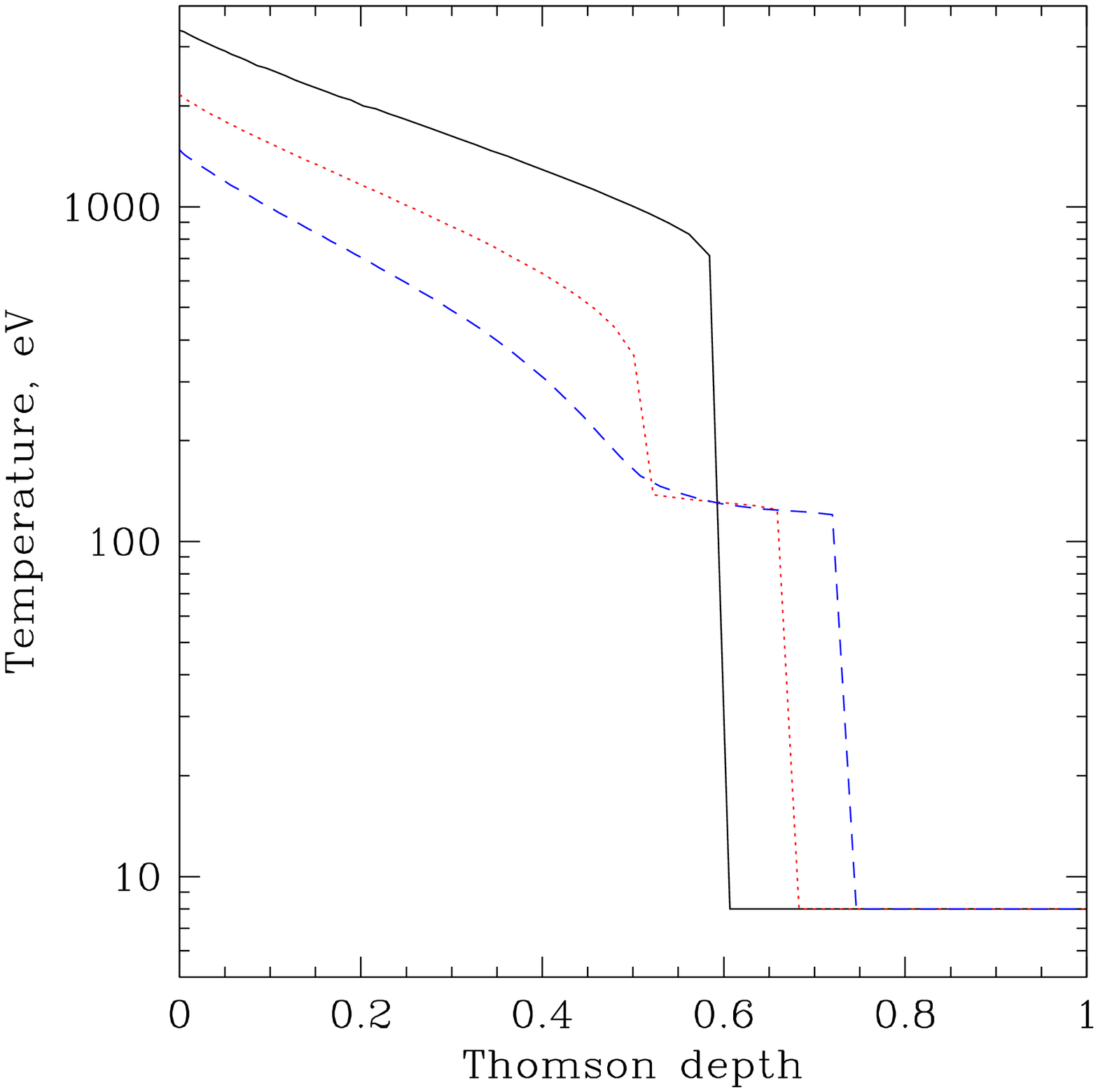,width=.45\textwidth,angle=0}}
\caption{Temperature profiles of the illuminated gas calculated for
the same geometry and total X-ray luminosity, $L_x$, etc., with only
the high-energy part of the illuminating spectrum varied between the
tests. The solid curve shows the temperature profile for the case of a
sharp cutoff at $E=200$ keV, while the dotted and the dashed ones show
tests with an exponential roll-over at $E_c = 150$ and $75$ keV,
respectively.}
\label{fig:temp_cut}
\end{figure*}

\begin{figure*}[H]
\centerline{\psfig{file=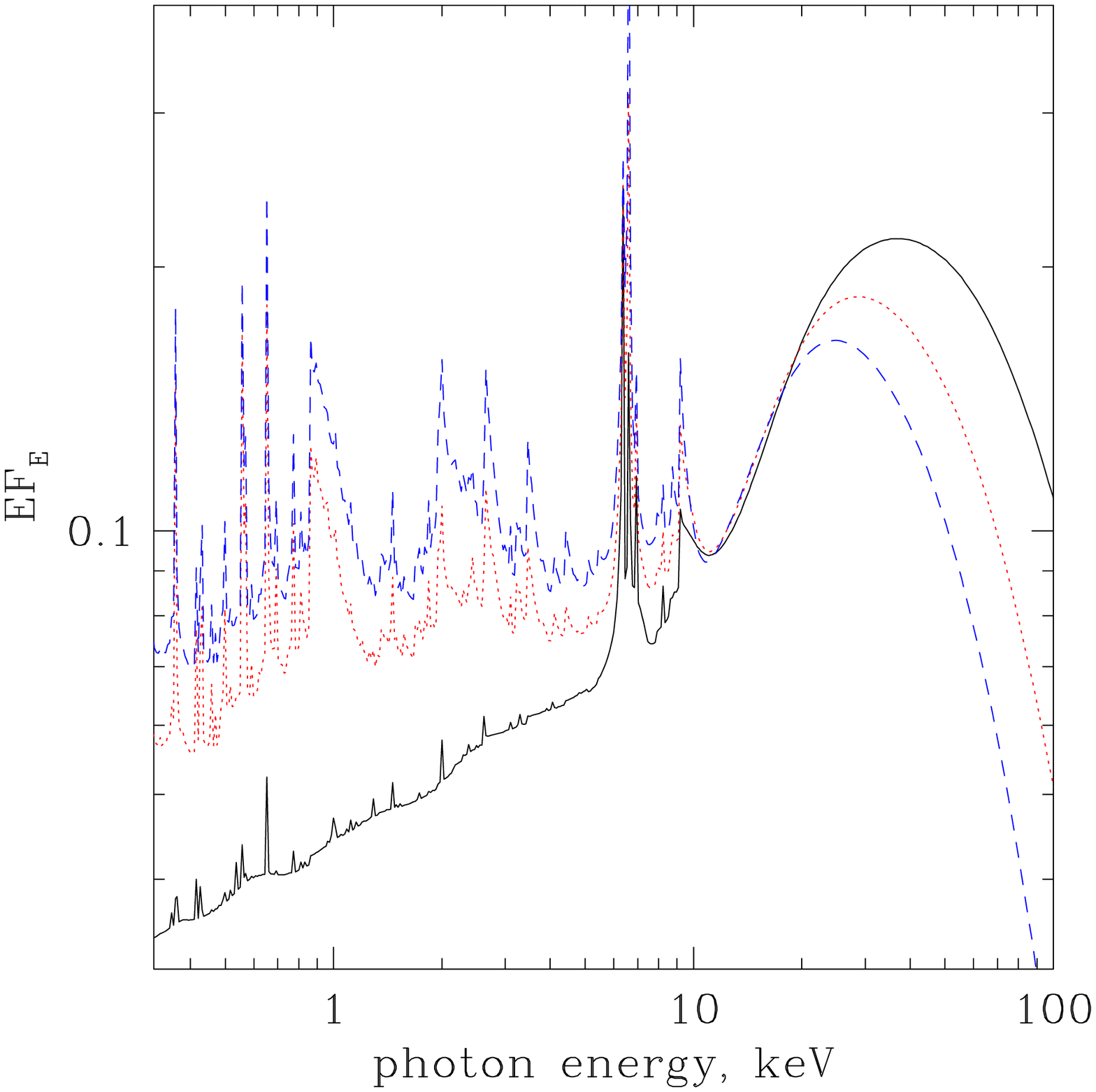,width=.55\textwidth,angle=0}}
\caption{Reflected spectra for the tests presented in
Fig. \ref{fig:temp_cut}. The meaning of the curves is the same as in
Fig. \ref{fig:temp_cut}.}
\label{fig:cut1}
\end{figure*}


\appendix

The importance of the high energy rollover in the spectrum was
discussed by Nayakshin \& Kallman (2001), but no quantitative tests
were presented. Therefore, using the code of NKK, we perform here
several tests that validate listing $E_c$ as a major unknown. For
simplicity, we choose the geometry of the lamppost model with $h_x = 6
R_S$ and $\dm = 0.01$. The illuminating X-ray spectrum is a power-law
with photon spectral index $\Gamma = 1.8$. 
Figure \ref{fig:temp_cut} shows the temperature profiles for the
illuminated gas for three different shapes of the high energy cutoff.
The solid curve shows the reflected spectrum for the case of a sharp
cutoff at $E=200$ keV, whereas the dotted and the dashed correspond to
tests with an exponential roll-over at $E_c = 150$ and $75$ keV,
respectively. The main effects in the evolution of the temperature
profiles, from the sharply cut to the $E_c = 75$ keV curve, seem to be

Note (1) the decrease in the gas temperature on the top of the skin;
(2) an appearance and increase in the extent of the
``mid-temperature'' step at $k T\sim 150 eV$ as $E_c$ decreases.
The physical mechanism behind the changes in the temperature profiles
and the reflected spectra is the change in the photo-ionization
equilibrium curve -- the so-called ``S-curve'' (see, e.g.,
Krolik et al. 1981; Fig. 1 in NKK).  As shown by
Nayakshin (2000a), the transition between the hot and the cold parts
of the ionized gas occurs at the gas pressure
\begin{equation}
\pgas \sim \pcrit = 0.032 T_1^{3/2}\, J/c\;,
\label{pcrit}
\end{equation}
where $T_1$ is the Compton temperature of the local radiation field,
and $J$ is the local radiation energy density. As the high energy
cuttoff $E_c$ decreases, $\pcrit$ decreases as well, and this is why
there is somewhat less of the ionized skin and it is also cooler for
smaller $E_c$\footnote{This is quite similar to changes in the S-curve
shown in Fig. 11 of Nayakshin \& Kallman (2001), even though
there these changes were caused by the increase of the ratio
$J_x/J_{bb}$ rather than $E_c$.}.

Figure \ref{fig:cut1} shows the reflected spectra at the inclination
angle $\mu = \cos i = 0.85$ ($\mu = 1$ corresponds to the normal to
the disk). The \fe line EW changes from about 82 eV to more than 200
eV. Clearly, the variations in the spectra induced by the changes in
the cutoff shape are large and non-trivial, and therefore the exact
value of $E_c$ is indeed important in the determination of the \fe
line emission from the disk.

{}

\end{document}